\documentstyle[preprint,aps,epsf,tighten]{revtex}	

\begin{document}
%
%
\preprint{$
\begin{array}{l}
\mbox{BA-02-33}\\
\mbox{FERMILAB-Pub-02/219-T}\\
\mbox{October 2002}\\[0.3in]
\end{array}
$}
\title{Lifting a Realistic SO(10) Grand Unified Model 
	to Five Dimensions}
\author{Carl H. Albright}
\address{Department of Physics, Northern Illinois University, DeKalb, IL
60115\\
       and\\
Fermi National Accelerator Laboratory, P.O. Box 500, Batavia, IL
60510\footnote{electronic address: albright@fnal.gov}}
\author{S.M. Barr}
\address{Bartol Research Institute,
University of Delaware, Newark, DE 19716\footnote{electronic address:
smbarr@bartol.udel.edu}}
\maketitle
\thispagestyle{empty}
\begin{abstract}

It has been shown recently that the problem of rapid proton decay induced
by dimension-five operators
arising from the exchange of colored Higgsinos can be simply avoided in 
grand unified models where a fifth spatial dimension is compactified
on an orbifold. Here we demonstrate that this idea can be used to
solve the Higgsino-mediated proton decay problem in any realistic
$SO(10)$ model by lifting that model to five dimensions. A particular 
$SO(10)$ model that has been proposed to explain
the pattern of quark and lepton masses and mixings is used as an example.
The idea is to break the $SO(10)$ down to the Pati-Salam symmetry by the 
orbifold boundary conditions. The entire 
four-dimensional $SO(10)$ model is placed on the physical $SO(10)$ brane 
except for the gauge fields, the ${\bf 45}$ and a single ${\bf 10}$   
of Higgs fields, which are placed in the five-dimensional bulk.  The 
structure of the Higgs superpotential can be somewhat simplified in doing 
so, while the Yukawa superpotential and mass matrices derived from it 
remain essentially unaltered.
\end{abstract}
%
\pacs{PACS numbers: 12.15Ff, 12.10.Dm, 12.60.Jv, 14.60.Pq}
%
%

\section{INTRODUCTION}

There are several pieces of evidence in favor of supersymmetric
grand unification (SUSY GUTs).
There is the impressive unification of the gauge couplings of the minimal 
supersymmetric standard model (MSSM) when extrapolated to high energies. 
There is the existence of neutrino mass whose magnitude corresponds very
well to what is expected from the see-saw mechanism when the 
right-handed neutrino mass is taken to be of order $M_{GUT}$, the scale
at which gauge couplings unify.
And there are various features of quark and lepton masses
and mixings that can be elegantly accounted for by grand unified 
symmetries. A well known example of this is the relation $m_b(M_{GUT}) = 
m_{\tau}(M_{GUT})$ \cite{mbmtau}.
Another is that the largeness of the atmospheric neutrino mixing angle 
may be related to the smallness of the corresponding quark mixing angle
$V_{cb}$ by grand unified symmetries in so-called
``lopsided'' models \cite{lopsided}. It is possible to construct GUT models 
of quark and lepton mass that are quite simple and predictive.

On the other hand, there are also serious problems with SUSY GUTs, 
in particular, the problem of doublet-triplet splitting and the closely 
related problem of dimension-five proton decay operators that arise from
the exchange of colored Higgsinos. In fact, given better limits on proton 
decay and better calculations of matrix elements recently, the problem 
of Higgsino-mediated proton decay has become quite 
severe \cite{pdkproblem}. Some of the more successful
GUT models of quark and lepton masses (at least in their original
forms) can no longer be claimed to be consistent with the improved 
experimental proton decay limits.

It is noteworthy that the successful
features of SUSY GUTs have to do with the gauge interactions and the quark
and lepton sectors, whereas all of their difficulties have to do with the Higgs
sector. This suggests that SUSY GUTs may require some radically new idea
for explaining how gauge symmetries break. It has been known since the
rebirth of Kaluza-Klein theories and superstring theory in the 1980's
that theories with extra dimensions allow interesting new ways of breaking
gauge symmetries and of solving the doublet-triplet splitting problem. 
Recently, it has been shown that in models with only one or two extra 
space dimensions orbifold compactification can break grand unified 
symmetries in such ways as to resolve rather simply both the doublet-triplet 
splitting and proton decay problems \cite{orbifold,DM}. These developments 
give good reason to suspect that extra dimensions may be the missing ingredient 
in the idea of grand unification.

The question naturally arises whether these ideas allow one to cure the
proton decay problem in actual realistic four-dimensional SUSY GUT models of 
quark and lepton masses. In this paper we show that they can. We consider a 
specific published $SO(10)$ model that is quite successful in 
reproducing the spectrum of quark and lepton masses and mixings --- including 
neutrino mixings --- with very few free parameters \cite{abb,ab,LMA}. 
We show that by embedding
this model in five-dimensions not only is the problem of proton decay
resolved, but the model is in several ways actually simplified in structure.

The paper is organized as follows. In Sec. II, we briefly review the 
four-dimensional $SO(10)$ model of quark and lepton masses, emphasizing in 
particular how the $SO(10)$ symmetry plays a role in explaining many features
of the data. We also review the Higgs sector that is needed in realistic 
$SO(10)$ models, discussing where the general problems lie 
and what must be done in four dimensions to resolve them. In Sec. III, we 
show how to embed the model described in Sec. II in five dimensions. We 
will see that to do this successfully the $SO(10)$ symmetry must be broken 
not only by orbifold compactification but also by a Higgs in the adjoint 
representation of $SO(10)$. We will see how this embedding in higher
dimension allows the Higgs sector to be significantly simplified.

\section{A REALISTIC FOUR-DIMENSIONAL SO(10) MODEL}
\subsection{The quark and lepton sector}

The model we shall study is that proposed in \cite{abb,ab,LMA}, 
which is of the ``lopsided'' type. The Dirac mass matrices of the quarks 
and leptons have the form

\begin{equation}
\begin{array}{ll}
M_U = \left( \begin{array}{ccc} \eta & 0 & 0 \\ 0 & 0 & \epsilon/3 \\
0 & - \epsilon/3 & 1 \end{array} \right) m_U, \;\;\; &
M_D = \left( \begin{array}{ccc} \eta & \delta & \delta' \\ \delta & 0 & 
\sigma + \epsilon/3 \\
\delta' & - \epsilon/3 & 1 \end{array} \right) m_D, \\ & \\ 
M_{\nu}^{Dirac} = \left( \begin{array}{ccc} \eta & 0 & 0 \\ 0 & 0 & -\epsilon \\
0 & \epsilon & 1 \end{array} \right) m_U, \;\;\; &
M_L = \left( \begin{array}{ccc} \eta & \delta & \delta' \\ \delta & 0 & 
-\epsilon \\
\delta' & \sigma + \epsilon & 1 \end{array} \right) m_D. \end{array}
\end{equation}

\noindent
The convention being used here is that the left-handed fermion fields
multiply the matrices from the right.
As explained in Ref. \cite{ab}, an excellent fit is obtained for the 
nine quark and charged lepton masses and four CKM mixing parameters with 
the seven real parameters and complex phase for $\delta'$ after evolution
downward from the GUT scale.  In addition, the atmospheric neutrino mixing 
angle is predicted to be very close to maximal, as observed.  A simple 
hierarchical form for the right-handed Majorana mass matrix in terms of 
three or four additional parameters leads to the large angle solar mixing 
solution (LMA) that is favored experimentally.
This is discussed in Refs. \cite{LMA,bimaximal}.

The terms in Eq. (1) come from six operators:

\begin{equation}
\begin{array}{lcl}
O_{33} & = & {\bf 16}_3 {\bf 16}_3 {\bf 10}_H, \\
& & \\
O_{23} & = & {\bf 16}_2 {\bf 16}_3 {\bf 10}_H {\bf 45}_H/M_{GUT}, \\
& & \\
O'_{23} & = & ({\bf 16}_2 {\bf 16}_H)({\bf 16}_3 {\bf 16}'_H)/M_{GUT}, \\
& & \\
O_{13} & = & {\bf 16}_1 {\bf 16}_3 {\bf 10}''_H, \\
& & \\
O_{12} & = & {\bf 16}_1 {\bf 16}_2 {\bf 10}'_H, \\
& & \\
O_{11} & = & {\bf 16}_1 {\bf 16}_1 {\bf 10}_H. 
\end{array}
\end{equation}

\noindent
The Higgs fields are denoted by the subscript `$H$'. The quark and lepton
multiplets, which are spinors of $SO(10)$, are distinguished by a family
label. The adjoint Higgs field ${\bf 45}_H$ is assumed to have a vacuum 
expectation value (VEV) proportional to the generator $B-L$. The
${\bf 10}_H$ has weak-scale VEVs in both the $(1,2, -\frac{1}{2})$ and
$(1,2, \frac{1}{2})$ components (under $SU(3)_c \times SU(2)_L \times U(1)_Y$),
while the ${\bf 10}'_H$ and ${\bf 10}''_H$ have VEVs only
in the $(1,2, -\frac{1}{2})$ component. The ${\bf 16}_H$ has a GUT-scale
VEV in the $(1,1,0)$ component, and ${\bf 16}'_H$ has a weak-scale VEV
in the $(1,2, -\frac{1}{2})$ component.

Many of the features of the mass matrices in Eq. (1) can be understood in
group-theoretical terms. The $\epsilon$ entries come from the operator
$O_{23}$ in Eq. (2). The factors of $\pm 1$ and $\pm \frac{1}{3}$ that
accompany these $\epsilon$ entries arise from the fact that 
$\langle {\bf 45}_H \rangle \propto B-L$. Moreover, it can be shown
that with $\langle {\bf 45}_H \rangle \propto B-L$ the $\epsilon$
entries must enter antisymmetrically in flavor, as in fact
they do in Eq. (1). (In $O_{23}$ there are two ways to contract the
fields to form an $SO(10)$ singlet. One of these contractions 
is flavor-symmetric, the other flavor-antisymmetric. The flavor-symmetric
piece vanishes if $\langle {\bf 45}_H \rangle \propto B-L$.)
The $\sigma$ entries in Eq. (1) come from the operator $O'_{23}$.
In the expression defining this operator in Eq. (2) the parentheses
mean that the fields inside the parentheses are contracted into a
${\bf 10}$ of $SO(10)$. From the fact that the VEV of ${\bf 16}_H$
points in the $SU(5)$-singlet direction, one sees that in
$SU(5)$ language $O'_{23} = {\bf 10}_2 \overline{{\bf 5}}_3 
\overline{{\bf 5}}_H$. Consequently, the $\sigma$ entries appear
in a ``lopsided'' way and are also transposed between $M_L$ and $M_D$.
In Eq. (1), the entries denoted by 1 come from $O_{33}$ and the entries 
denoted by $\delta$ and $\delta'$ come from $O_{12}$ and $O_{13}$, 
respectively.  All of these are flavor-symmetric since ${\bf 10}$ is in the 
symmetric product of ${\bf 16} \times {\bf 16}$. 

Four of the operators in Eq. (2) are non-renormalizable. These can arise
from integration out of real representations of quarks and leptons.
For example the operator ${\bf 16}_2 {\bf 16}_3 {\bf 10}_H {\bf 45}_H/M_{GUT}$
can come from integrating out a vector-like family-antifamily pair,
${\bf 16} + \overline{{\bf 16}}$, having the renormalizable couplings
${\bf 16}_3 \overline{{\bf 16}} \; {\bf 45}_H + {\bf 16}_2 {\bf 16} \; 
{\bf 10}_H + M_{16} \overline{{\bf 16}} \; {\bf 16}$. The operator
$({\bf 16}_2 {\bf 16}_H)({\bf 16}_3 {\bf 16}'_H)/M_{GUT}$ can come from 
integrating out a ${\bf 10}$ of quarks and leptons having the 
renormalizable couplings ${\bf 16}_3 {\bf 10} \; {\bf 16}'_H + 
{\bf 16}_2 {\bf 10} \; {\bf 16}_H + M_{10} {\bf 10} \; {\bf 10}$.  The operator
$O_{11}$ involves the integration out of additional vector-like fermions,
while the operator $O_{12}$ involves the integration out of other massive
Higgs fields, which we do not spell out here.  Aside from these last two
operators, those listed in Eq. (2) can arise from the following
renormalizable terms in the superpotential involving the
quark and lepton fields:

\begin{equation}
\begin{array}{ccl}
W_{Yukawa} & = & {\bf 16}_3 {\bf 16}_3 {\bf 10}_H + 
  {\bf 16}_3 \overline{{\bf 16}} \; {\bf 45}_H + {\bf 16}_2 {\bf 16} \; 
  {\bf 10}_H + M_{16} \overline{{\bf 16}} \; {\bf 16} \\ & & \\
& + &  {\bf 16}_3 {\bf 10} \; {\bf 16}'_H
+ {\bf 16}_2 {\bf 10} \; {\bf 16}_H + M_{10} {\bf 10} \; {\bf 10} \\ & & \\
& + & {\bf 16}_1 {\bf 16}_3 {\bf 10}''_H. \\
\end{array}
\end{equation}

The field content of this model consists of several vectors, spinors, 
and antispinors of $SO(10)$ together with the adjoint of gauge fields
and a single adjoint of Higgs. This adjoint Higgs field plays three very 
crucial roles. First, it participates in the breaking of $SO(10)$ down to 
the Standard Model group. Second, it produces the doublet-triplet splitting. 
To do this it is crucial that its VEV 
is proportional to $B-L$, since the triplet Higgs that must acquire superheavy 
masses have $B-L \neq 0$, whereas the doublets that must not get such 
masses have $B-L = 0$. And, third, the adjoint Higgs, by coupling to
the quarks and leptons, introduces the breaking of $SO(10)$ into the
quark and lepton masses, so that the ``bad'' relations $m_s = m_{\mu}$
and $m_d = m_e$ can be avoided. More precisely, it introduces the
factors of 1/3 into those matrices that are responsible for giving
the well-known and successful
Georgi-Jarlskog relations $m_{\mu}(M_{GUT}) \cong 3 m_s(M_{GUT})$ and
$m_e(M_{GUT}) \cong m_d(M_{GUT})/3$ \cite{gj}. These factors of 1/3 also 
require that the VEV of ${\bf 45}_H$ be proportional to $B-L$.

\subsection{The Higgs sector}

Let us review the Higgs structure that is needed to break $SO(10)$ in 
four dimensions, both in general and in this model specifically.
In general, there are at least four sectors that are needed in the Higgs
superpotential: the doublet-triplet-splitting sector, the adjoint sector,
the spinor sector, and the adjoint-spinor-coupling sector \cite{bb93}. 

The doublet-triplet splitting in $SO(10)$ models in four dimensions
must be done by the Dimopoulos-Wilczek mechanism \cite{dw,bb93}. The simplest 
form of this mechanism assumes the existence of a term ${\bf 10}_H
{\bf 45}_H \tilde{{\bf 10}}_H$. 
(Throughout this paper we will not bother
 to write the dimensionless coefficients of terms in the superpotential.)
There must be two distinct ${\bf 10}$s in this term 
because the adjoint is in the antisymmetric product of 
${\bf 10} \times {\bf 10}$. If $\langle {\bf 45}_H \rangle \propto
B-L$, then, for reasons already explained, only the color-triplet
fields in the vector Higgs multiplets obtain mass from this term,
while the weak doublets all remain massless. Since there are four
doublets all told in ${\bf 10}_H + \tilde{{\bf 10}}_H$ half of these
must be made superheavy to reproduce the MSSM. This can be achieved
by the simple term $M (\tilde{{\bf 10}}_H)^2$. Thus the 
doublet-triplet-splitting sector has the terms
\begin{equation}
W_{2/3} = {\bf 10}_H {\bf 45}_H \tilde{{\bf 10}}_H
+ M_1 (\tilde{{\bf 10}}_H)^2.\\
\end{equation}

The adjoint sector of the superpotential is responsible for forcing
the adjoint Higgs to have a VEV in the $B-L$ direction. The simplest
possibility is
\begin{equation}
W_{45} = {\rm tr} ({\bf 45}_H)^4/M_2 - M_3 {\rm tr} ({\bf 45}_H)^2.
\end{equation}

\noindent
If $\langle {\bf 45}_H \rangle$ has the form 
${\rm diag}(a_1, a_2, a_3, a_4, a_5) \otimes i \tau_2$, then the
superpotential can be written $W_{45} = \sum_i (a_i^4/M_2 - M_3 a_i^2)$.
Clearly, the scalar potential is minimized by $a_i = 0$ or $(M_2 M_3/2)^{1/2}$.
One solution is $\langle {\bf 45}_H \rangle = (M_2 M_3/2)^{1/2}$ 
diag$(0,0,1,1,1,) \times i \tau_2 = \frac{3}{2} (M_2 M_3/2)^{1/2} (B-L)$.
The quartic term could be a Planck-scale effect, but then in order
to make the vacuum expectation value of the adjoint be of order
$M_{GUT}$, the parameter $M_3$ would have to be $O(M_{GUT}^2/M_{Pl})$.
On the other hand, if the quartic term arises
from integrating out some field whose mass is of order $M_{GUT}$, then 
that field would have to be a ${\bf 54}$
or some larger representation. A singlet would not do, since integrating
out a singlet could yield a quartic term in ${\bf 45}_H$ only of
the form $({\rm tr}({\bf 45}_H)^2)^2/M_2$. This would leave a continuous
degeneracy of all minima satisfying $\sum_i a_i^2 = M_2 M_3/2$.
Obtaining a satisfactory quartic term for the adjoint is thus a 
significant difficulty for such $SO(10)$ models in four dimensions.
We shall see that a solution to this is quite simple in five-dimensional
models.

The spinor sector is required to generate GUT-scale VEVs for the spinor 
Higgs fields ${\bf 16}_H + \overline{{\bf 16}}_H$. These VEVs break
$SO(10)$ down to an $SU(5)$ subgroup (this could alternatively
be done by $\overline{{\bf 126}}_H + {\bf 126}_H$), and the adjoint VEV 
proportional to $B-L$ further breaks it down to the Standard Model group
$SU(3)_c\times SU(2)_L \times U(1)_Y$. 
It is very easy to construct a
superpotential to give the spinors such VEVs. The simplest is

\begin{equation}
W_{16} = (\overline{{\bf 16}}_H {\bf 16}_H - M_4^2) {\bf 1}_{1H}.
\end{equation}

As explained in
Ref. \cite{bb93} there must be a sector that couples the adjoint Higgs to the
spinor Higgs. If there is none, then there is nothing to determine the
relative orientation of the adjoint and spinor VEVs. Corresponding to
this degeneracy there would be unwanted goldstone bosons. This is
a serious difficulty in $SO(10)$ model building in four dimensions
since most ways to couple the ${\bf 45}_H$ to the $\overline{{\bf 16}}_H
+ {\bf 16}_H$ (for example, by the obvious term $\overline{{\bf 16}}_H 
{\bf 45}_H {\bf 16}_H$) would destabilize the adjoint VEV so that it would
no longer exactly satisfy the Dimopoulos-Wilczek condition 
$\langle {\bf 45}_H \rangle \propto B-L$. As a result, the doublet Higgs
fields in ${\bf 10}_H$ would acquire GUT-scale mass. One way around
this dilemma was suggested in \cite{bb93} where it was pointed out that
if there are three distinct adjoint fields the adjoint-spinor coupling can
be done without destabilizing the Dimopoulos-Wilczek form of the adjoint
VEV. In Ref. \cite{br} it was then shown that it could be done with only
a single adjoint Higgs field, but with additional spinor Higgs fields,
if the adjoint-spinor couplings in the superpotential have the form

\begin{equation}
W_{45-16} = \overline{{\bf 16}}_H( {\bf 45}_H - {\bf 1}_{2H}) {\bf 16}'_H
+ \overline{{\bf 16}}'_H ({\bf 45}_H - {\bf 1}_{3H}) {\bf 16}_H.
\end{equation}

\noindent
The primed spinor fields here do not acquire VEVs (at least at the GUT
scale, they may at the weak scale). Consequently,
the $F$ terms for the adjoint Higgs field and the unprimed spinor Higgs
fields do not get contributions from the terms in $W_{45-16}$.
The fields ${\bf 1}_{2H}$ and ${\bf 1}_{3H}$ are 
``sliding singlets,'' whose VEVs are free
to adjust to make the $F$ terms of the primed spinor Higgs fields vanish.

The four sectors just discussed are the only ones that are 
necessary in four-dimensional models based on $SO(10)$ to break
$SO(10)$ all the way down to the Standard Model group. However, 
additional sectors may be necessary for other reasons. For example,
in the model of quark and lepton masses we are discussing, as well
as in some other published models, it is assumed that there are
spinor Higgs that have only $SU(2)_L \times U(1)_Y$-breaking, 
weak-scale VEVs. The most economical possibility
would be for those Higgs to be the primed spinors that appear in 
$W_{45-16}$. Then to induce the weak-scale VEV we want, one must 
have terms that mix these primed spinors with the ${\bf 10}_H$.
One possibility is $\overline{{\bf 16}}_H \overline{{\bf 16}}_H {\bf 10}_H$.
Then the term $F_{\overline{16}_H}^* F_{\overline{16}_H}$ mixes
the $(1,2, -\frac{1}{2})$ components of ${\bf 10}_H$ and ${\bf 16}'_H$.
In this model the ${\bf 10}'_H$ is assumed to have a weak-scale VEV
in the $(1,2,-\frac{1}{2})$ direction, but not the $(1,2,+\frac{1}{2})$
direction. This can be achieved by introducing terms ${\bf 10}'_H 
{\bf 16}_H {\bf 16}'_H + M_5 {\bf 10}'_H {\bf 10}'_H$.
Then the VEV of ${\bf 16}'_H$ in the $(1,2,-\frac{1}{2})$ direction
induces one in ${\bf 10}'_H$. In this model, then, there is a sector
of the superpotential that mixes the vector and spinor Higgs, having the terms

\begin{equation}
W_{10-16} = \overline{{\bf 16}}_H \overline{{\bf 16}}_H {\bf 10}_{H} +
{\bf 10}'_H 
{\bf 16}_H {\bf 16}'_H + M_5 {\bf 10}'_H {\bf 10}'_H.
\end{equation}

A very important consideration in four-dimensional models based on $SO(10)$
is preserving the gauge hierarchy in a natural way. There are two classes 
of terms that can destroy the gauge hierarchy. Class A consists of
operators of the type $M({\bf 10}_H)^2$, $M {\bf 10}_H \tilde{{\bf 10}}_H$, 
$({\bf 10}_H)^2 ({\bf 45}_H)^2/M_{Pl}$, 
${\bf 10}_H \tilde{{\bf 10}}_H ({\bf 45}_H)^2/M_{Pl}$,   
$({\bf 10}_H)^2 (\overline{{\bf 16}}_H {\bf 16}_H)/M_{Pl}$,
etc., which directly produce too large a mass for the MSSM doublets
$H_u$ and $H_d$. Such a mass must not be larger than order $M_{GUT}^5/M_{Pl}^4$.
Class B consists of operators that destabilize the
Dimopoulos-Wilczek form of the VEV of the adjoint Higgs field. If one writes
$\langle {\bf 45}_H \rangle = A (B-L) + B I_{3R}$, then $B/A$ must be
less than or about $10^{-13} \sim (M_{GUT}/M_{Pl})^4$. Many terms that
directly couple the adjoint Higgs to the spinor Higgs, 
such as $\overline{{\bf 16}}_H {\bf 45}_H {\bf 16}_H$
and $\overline{{\bf 16}}_H ({\bf 45}_H)^2 {\bf 16}_H/M_{Pl}$, fall 
into this class.

Such operators may be forbidden by a flavor symmetry. For example, in 
\cite{ab} a $U(1) \times Z_2 \times Z_2$ flavor symmetry was
shown to protect the gauge hierarchy. The price of such a symmetry is that
the field content of the model and the form of the couplings must be
somewhat more complicated. For example, the terms in Eqs. (5) and (6) are
too simple as they stand. Eq. (5) would imply that $({\bf 45}_H)^2$ is
neutral under all symmetries, and Eq. (6) would imply the same thing about
$\overline{{\bf 16}}_H {\bf 16}_H$. Consequently the term 
$\overline{{\bf 16}}_H ({\bf 45}_H)^2 {\bf 16}_H/M_{Pl}$, which would
destroy the gauge hierarchy, would be allowed by all symmetries.
In \cite{ab} this problem was avoided by replacing
the expression in Eq. (6) by one of the form
$((\overline{{\bf 16}}_H {\bf 16}_H)^2/M^2 - M^2) {\bf 1}_H$. This
allows $\overline{{\bf 16}}_H {\bf 16}_H$ and thus 
$\overline{{\bf 16}}_H ({\bf 45}_H)^2 {\bf 16}_H/M_{Pl}$
to be odd under a $Z_2$. In order to impose the abelian flavor symmetry
other complications had to be introduced in \cite{ab}.
These included (a) replacing most of the dimensionful parameters that appear
in Eqs. (3) - (7), and some of the dimensionless couplings, by the VEVs 
of gauge-singlet Higgs fields, (b) having two ${\bf 10}$s of quarks and 
leptons instead of just one appearing in the terms in Eq. (3) that give rise
to the effective operator $O'_{23}$, (c) having different
${\bf 10}'_H$ fields appearing in the operators $O_{12}$ and $O_{13}$, and
(d) obtaining the operator $O_{11}$ by integrating out an additional
spinor-antispinor pair of quarks and leptons. 

We will see in the next section that one of the great advantages 
of embedding the $SO(10)$ model in five-dimensions is that it is
no longer necessary to have the Dimopoulos-Wilczek form be highly exact,
since the doublet-triplet splitting is produced by the orbifold
compactification rather than by the adjoint Higgs field. This means that
it is no longer necessary to worry about operators of Class B. 

We will not further discuss the details of the model of quark and lepton
masses proposed in Refs. \cite{abb,ab,LMA}. Those interested in them
can consult those papers. We now turn to the question of whether and how
this model can be embedded in five space-time dimensions. The reason
for doing so is that as a four-dimensional model it potentially has a serious
difficulty with Higgsino-mediated proton decay. To understand why
going to five dimensions resolves the proton-decay problem it is 
useful to review why there is a problem in four dimensions.
The Higgs doublets $H_d$ and $H_u$ of the MSSM sit in the grand unified 
multiplets as follows: $(H_d, H_{\overline{3}}) = 
\overline{{\bf 5}}_H \subset {\bf 10}_H$, and $(H_u, H_{3}) =
{\bf 5}_H \subset {\bf 10}_H$. The doublets $H_d$ and $H_u$ 
do not get mass from the terms in Eq. (4). However, their color-triplet 
partners, $H_{\overline{3}}$ and $H_{3}$, do get
superheavy masses with the color triplets $\tilde{H}_{3}$ and
$\tilde{H}_{\overline{3}}$ in $\tilde{{\bf 10}}_H$ from
the first term in Eq. (4). In turn, $\tilde{H}_{3}$ and 
$\tilde{H}_{\overline{3}}$ get mass with each other from the second 
term in Eq. (4). Consequently, if the triplets in $\tilde{{\bf 10}}_H$ are 
integrated out, one has effectively a mass term linking $H_{\overline{3}}$ to 
$H_{3}$. It is this that allows a diagram in which exchange of color-triplet
Higgsinos mediates proton decay. 

A crucial point in the five-dimensional model that we shall
outline in the next section is that the color-triplet Higgsinos
$H_{\overline{3}}$ and $H_{3}$ can get superlarge mass without
having a mass term that links them to each other. This happens because
they each get a Kaluza-Klein mass that links them not to each other
but to Kaluza-Klein modes in the same five-dimensional hyperfield.

\section{LIFTING THE MODEL TO FIVE DIMENSIONS}

In the four-dimensional model the Higgs field in the adjoint representation
played three crucial roles: it helped break $SO(10)$ down to the Standard
Model group, it did the doublet-triplet splitting, and it gave the
Clebsch factors of $1/3$ in the quark mass matrices that allowed
the model to reproduce the Georgi-Jarlskog relations between quark and
lepton masses. In five-dimensional models it has been shown that
the first two tasks can be accomplished by orbifold
compactification rather than by an adjoint Higgs. It would be exceedingly
interesting, therefore, to find a model in which orbifold compactification
could also account for the Georgi-Jarlskog factors. In that case, the
adjoint Higgs could be dispensed with altogether. However, it is not
easy to see how the desired Clebsch factors of $1/3$ can be predicted
(rather than merely being accommodated) by 
orbifold compactification. Therefore, it seems that in extending the
model that we have been examining to higher dimension it is necessary to
retain an adjoint Higgs field. 

In constructing the five-dimensional version of the model we follow 
the procedure explained in \cite{orbifold,DM}. We suppose that the fifth 
dimension is compactified on a $S^1/(Z_2 \times Z'_2)$
orbifold. The $S^1$ is defined by $y \equiv y + 2 n \pi R$; the $Z_2$ maps 
$y \leftrightarrow -y$ and the $Z'_2$ maps $y' \leftrightarrow - y'$, 
where $y' = y + \pi R/2$. The fundamental region may therefore be taken 
to be $-\pi R/2 \leq y \leq 0$. Point $O$ at $y=0$ (the fixed point of $Z_2$)
is the ``visible brane,'' while point $O'$ at $y'=0$ (the fixed point of 
$Z'_2$) is the ``hidden brane''.
The compactification scale $1/R \equiv M_C$
is assumed to be close to the scale at which the gauge couplings
unify, i.e., the GUT scale where $M_{GUT} \sim 10^{16}$ GeV.

The generic bulk field $\phi(x^\mu,y)$, where $\mu=0,1,2,3$, has definite
parity assignment under $Z_2 \times Z'_2$ symmetry. Taking $P=\pm 1$ to be
the parity eigenvalue of the field $\phi(x^\mu,y)$ under $Z_2$ transformation
and $P'=\pm 1$ to be the parity eigenvalue under the $Z'_2$ transformation,
a field with $(P,P')=(\pm,\pm)$ can be denoted
$\phi^{PP'}(x^\mu,y)=\phi^{\pm \pm}(x^\mu,y)$. The Fourier series
expansion of the fields $\phi^{\pm \pm}(x^\mu,y)$ yields

\begin{equation}
\begin{array}{lcl}
\phi^{++}(x^\mu,y) & = & \frac{1}{\sqrt{2^{\delta_{n0}}\pi R}}
\sum^{\infty}_{n=0}
\phi^{++(2n)}(x^\mu) \cos \frac{2ny}{R},\\ & & \\
\phi^{+-}(x^\mu,y) & = & \frac{1}{\sqrt{\pi R}} \sum^{\infty}_{n=0}
\phi^{+-(2n+1)}(x^\mu) \cos \frac{(2n+1)y}{R},\\ & & \\
\phi^{-+}(x^\mu,y) & = & \frac{1}{\sqrt{\pi R}} \sum^{\infty}_{n=0}
\phi^{-+(2n+1)}(x^\mu) \sin \frac{(2n+1)y}{R},\\ & & \\
\phi^{--}(x^\mu,y) & = & \frac{1}{\sqrt{\pi R}} \sum^{\infty}_{n=0}
\phi^{--(2n+2)}(x^\mu) \sin \frac{(2n+2)y}{R}.
\end{array}
\end{equation}

In the effective theory in four dimensions all of the bulk fields 
have masses of order $M_C$ except the Kaluza-Klein zero
mode $\phi^{++(0)}$ of $\phi^{++}(x^\mu,y)$, which remains massless.
Moreover, fields $\phi^{-\pm}(x^\mu,y)$ vanish on the visible brane and
fields $\phi^{\pm-}(x^\mu,y)$ vanish on the hidden brane.

In our model, we assume that gauge fields and two multiplets of Higgs fields,
${\bf 10}_H$ and ${\bf 45}_H$, exist in the 
five-dimensional bulk, while the quark and lepton fields and the remaining 
Higgs fields exist on the visible brane at $O$.

The gauge fields in the bulk are of course in a vector supermultiplet of 
five-dimensional supersymmetry that is an
adjoint representation of $SO(10)$. We will denote it by ${\bf 45}_g$,
where the subscript `$g$' stands for `gauge'.
This vector supermultiplet decomposes
into a vector multiplet $V$ and a chiral multiplet $\Sigma$
of $N = 1$ supersymmetry in four dimensions.
The bulk Higgs fields ${\bf 10}_H$ and ${\bf 45}_H$ are 
in hypermultiplets of five-dimensional supersymmetry. Each hypermultiplet 
splits into two left-handed chiral
multiplets $\Phi$ and $\Phi^c$, having opposite gauge quantum numbers.

Following Derm\'{i}\v{s}ek and Mafi \cite{DM}, we will assume 
that the orbifold compactification breaks $SO(10)$ down to
the Pati-Salam group $G_{PS} = SU(4)_c \times SU(2)_L \times SU(2)_R$,
and that the further breaking to the Standard Model group is accomplished by
the spinor Higgs fields ${\bf 16}_H + \overline{{\bf 16}}_H$ that live on 
the visible brane through the ordinary four-dimensional Higgs mechanism.  
Under $G_{PS}$ the $SO(10)$ representations decompose as follows:
${\bf 45} \rightarrow (15,1,1)+ (1,3,1)+ (1,1,3) + (6,2,2)$; 
${\bf 10} \rightarrow (6,1,1) + (1,2,2)$;
${\bf 16} \rightarrow (4,2,1) + (\overline{4}, 1,2)$;
and $\overline{{\bf 16}} \rightarrow (\overline{4}, 2,1) + (4,1,2)$.
With these facts in mind we shall now discuss the
transformation of the various fields under the $Z_2 \times Z'_2$ parity
transformations.

The first $Z_2$ symmetry (the one we denote as unprimed)
is used to break supersymmetry to $N =1$ in four-dimensions.
($N=1$ in five dimensions is equivalent to $N=2$ in
four dimensions; so we are breaking half the supersymmetries.) To do this
we assume that under $Z_2$ the $V$ is even, $\Sigma$ is odd,
$\Phi$ are even, and $\Phi^c$ are odd. The $Z_2'$ is used to break
$SO(10)$ down to $G_{PS}$. The $(15,1,1)$, $(1,3,1)$ and $(1,1,3)$ of
${\bf 45}_g$ and ${\bf 45}_H$ are taken to be even under $Z_2'$, while 
the $(6,2,2)$ is taken to be odd. In ${\bf 10}_H$
the $(1,2,2)$ is taken to be even and the
$(6,1,1)$ odd.

All told we have

\begin{equation}
\begin{array}{ccl}
{\bf 45}_g & = & V^{++}_{(15,1,1)} + V^{++}_{(1,3,1)}+ V^{++}_{(1,1,3)} +
V^{+-}_{(6,2,2)} \\ & & \\ & + & 
\Sigma^{-+}_{(15,1,1)} + \Sigma^{-+}_{(1,3,1)} +
\Sigma^{-+}_{(1,1,3)} + \Sigma^{--}_{(6,2,2)}  \\ & & \\  
{\bf 45}_H & = & \Phi^{++}_{(15,1,1)} + \Phi^{++}_{(1,3,1))} +
\Phi^{++}_{(1,1,3)} + \Phi^{+-}_{(6,2,2)} \\ & & \\ &  
+ & \Phi^{c--}_{(15,1,1)} + \Phi^{c--}_{(1,3,1))} +
\Phi^{c--}_{(1,1,3)} +
\Phi^{c-+}_{(6,2,2)} \\ & & \\
{\bf 10}_H & = & \Phi^{++}_{(1,2,2)} +
\Phi^{+-}_{(6,1,1)} + \Phi^{c --}_{(1,2,2)} + \Phi^{c-+}_{(6,1,1)} \\
\end{array}
\end{equation}

\noindent
Massless zero modes of the Kaluza-Klein towers
exist only for fields with $Z_2 \times Z_2'$ parity $++$. 
The $\Phi^{++}_{(1,2,2)}$ in the ${\bf 10}_H$ contains the two light
doublets of the MSSM. Their color-triplet partners are made superheavy
by the orbifold compactification. The $\Phi^{++}_{(15,1,1)}$, 
$\Phi^{++}_{(1,3,1))}$, and $\Phi^{++}_{(1,1,3)}$ in the ${\bf 45}_H$
contain zero modes that do not get mass from the compactification.
However, these will obtain GUT-scale mass from minimizing the superpotential
in the four-dimensional effective theory or will be eaten by
gauge bosons and get GUT-scale mass that way. 

Having done with the parity assignment for the bulk fields we can turn our
attention to the brane physics. On the brane at $O$ we put all of the quark 
and lepton fields, i.e., not only the three families of spinors
${\bf 16}_i$ but also the real representations of quarks and leptons
that are integrated out to produce the effective operators of the form 
shown in Eq. (2). This includes the fields shown in Eq. (3).
We also place on the brane at $O$ all the Higgs fields,
except ${\bf 10}_H$ and ${\bf 45}_H$, which live in the bulk.

The $Z_2$ parity of fields that live on the visible brane, 
(i.e. ${\bf 16}_i$, $\overline{{\bf 16}}$, ${\bf 16}$, ${\bf 10}$,
$\overline{{\bf 16}}_H$, ${\bf 16}_H$, $\overline{{\bf 16}}'_H$, 
${\bf 16}'_H$, and ${\bf 1}_{aH}$) must be positive. 
The $Z'_2$ parity assignments must be consistent with those of the
components of ${\bf 45}_g$. For example, the quarks and leptons of the third
family must have parities ${\bf 16}_3 \longrightarrow (4,2,1)_3^{+ \pm}
+ (\overline{4}, 1,2)_3^{+ \mp}$. The superpotential on the visible
brane can have couplings between brane fields and bulk fields. 
For example, we assume there to be a Yukawa term of the form
${\bf 16}_3 {\bf 16}_3 {\bf 10}_H$. This expression is shorthand for

\begin{equation}
\begin{array}{ccl}
S_{16_3 16_3 10_H} & = & \int{{\rm d}^5x} \, 
\frac{1}{2} \left[\delta(y) - \delta(y-\pi R)\right] \sqrt{2 \pi R} 
\; \int{{\rm d}^2\theta} \, (4,2,1)_3 (\overline{4},1,2)_3 \Phi_{(1,2,2)}^{++}
\\
& & \\
& + &  \int{{\rm d}^5x} \, 
\frac{1}{2} \left[\delta(y) - \delta(y-\pi R)\right] \sqrt{2 \pi R} 
\; \int{{\rm d}^2\theta} \, (4,2,1)_3 (4,2,1)_3 \Phi_{(6,1,1)}^{+-} 
\\ & & \\ & + & {\rm h.c.}
\end{array}
\end{equation}

\noindent
Note that the products $(4,2,1)^{+ \pm} (\overline{4},1,2)^{+ \mp}
\Phi_{(1,2,2)}^{++}$ and $(4,2,1)^{+ \pm} (4,2,1)^{+ \pm} \Phi_{(6,1,1)}^{+-}$
are odd under $Z'_2$ which accounts for the minus sign difference between the 
two delta functions in both terms, since $Z'_2$ maps $y = 0$ onto $y = \pi R$. 
Even though the $\Phi_{(6,1,1)}$ gets mass of order $M_C$ from the 
compactification, it couples to the quarks and leptons on the visible brane.
None of the symmetries discussed so far would prevent a term of the form 
$\Phi_{(6,1,1)} \Phi_{(6,1,1)}$ in the superpotential on the visible
brane. If such a term were present, it would cause proton decay mediated
by the fermionic colored fields in the $\Phi_{(6,1,1)}$. 
However, such a mass term will be prevented from occurring by the abelian 
flavor symmetry to be discussed
later. (In an ordinary four-dimensional theory, such a mass term
or its equivalent is required to make the color triplet Higgs and Higgsinos
superheavy; in five-dimensional theories such a mass term is not needed
as the triplets get Kaluza-Klein masses from the compactification.)

Using the same shorthand notation, we can state the other couplings of the
superpotential on the visible brane. These can be divided into the Yukawa
terms of the quarks and leptons and the Higgs self-couplings.
The Yukawa terms can be chosen to have the same form
as in the four-dimensional version of the model, which was described in
the last section. However, the Higgs part of the superpotential
can be significantly simplified as a result of the embedding in
five dimensions. We saw in the last section that the Higgs superpotential 
in a four-dimensional $SO(10)$ model has at least four pieces: $W_{Higgs} =
W_{2/3} + W_{45} +  W_{45-16} + W_{16}$. Let us consider these
four pieces in turn. 

{\it The doublet-triplet splitting sector.}
In the four-dimensional version of the model, the doublet-triplet-splitting 
required the existence of two vector Higgs, which
were denoted ${\bf 10}_H$ and $\tilde{{\bf 10}}_H$, with couplings
of the form shown in Eq. (4). However, in the five-dimensional model
the doublet-triplet splitting is achieved by the orbifold compactification.
Consequently, there is no need for the field $\tilde{{\bf 10}}_H$ at all,
and no need for the terms in Eq. (4). That is, the entire
doublet-triplet piece of the superpotential $W_{2/3}$ can be
dispensed with. 

{\it The adjoint sector.} In the four-dimensional version 
of the model it was necessary, in order to obtain the Dimopoulos-Wilczek
form of the adjoint VEV, $\langle {\bf 45}_H \rangle \propto B-L$,
to have a non-renormalizable term of the form ${\rm tr} ({\bf 45}_H)^4/M$,
as shown in Eq. (5). As noted there, if this term arises as an effective
operator from integrating out a superheavy field, that field must be
a ${\bf 54}$ or something larger, whereas
if it does not come from integrating out some field, then one would
expect the denominator $M$ to be of order $M_{Pl}$. 
In five-dimensions, the problem of 
obtaining an adjoint VEV in the $B-L$ direction has a simple
solution that involves the physics on the ``hidden brane'' at $O'$.

The gauge symmetry on the hidden brane is the Pati-Salam group
$G_{PS} = SU(4)_c \times SU(2)_L \times SU(2)_R$. The components of the
adjoint Higgs field ${\bf 45}_H$ that are non-vanishing on the
hidden brane are those with $Z'_2 = +$, namely the $\Phi_{(15, 1, 1)}$,
$\Phi_{(1,3,1)}$, and $\Phi_{(1,1,3)}$ of $G_{PS}$, as shown in Eq. (10). 
This means that there is a superpotential on the hidden brane involving 
these components:

\begin{equation}
W_{O'} = {\rm tr}\ \Phi^3_{(15,1,1,)} - M {\rm tr}\ \Phi^2_{(15,1,1)} 
- M' {\rm tr}\ \Phi^2_{(1,3,1)} - M^{\prime \prime} {\rm tr}\ \Phi^2_{(1,1,3)}.
\end{equation}

\noindent
Note that a cubic term is allowed for $\Phi_{(15,1,1)}$, since in 
$SU(4)_c = SO(6)$ the totally symmetrized product of $15^3$
contains the singlet, whereas that is not true in $SO(10)$.
That is why in the four-dimensional $SO(10)$ model with only one adjoint 
field there had to be a term {\it quartic} in the adjoint field.
In $SO(10)$, only if there are three distinct adjoint fields can one
write down a cubic term for them. (The difference between $SO(10)$ and
$SO(6)$ in this regard is that in $SO(6)$ there is a rank six
antisymmetric tensor that allows one to write $\epsilon_{abcdef}
T^{[ab]} T^{[cd]} T^{[ef]}$.) This superpotential gives a scalar potential
one of whose (supersymmetric) minima is just $\langle \Phi_{(15,1,1)} \rangle
\propto B-L$, since of course $B-L$ is just one of the generators
of $SU(4)_c$. 

It is interesting that in the five-dimensional version of the model
one to some extent explains why the adjoint VEV prefers to point in 
the $B-L$ direction (as is useful in explaining the pattern of quark 
and lepton masses, as we saw in the last section). In five-dimensions 
the splitting of the doublet and triplet Higgs requires that the orbifold
compactification break $SO(10)$ down to the Pati-Salam group. And
the resulting structure of the superpotential has a term coming from the 
hidden brane that can drive breaking in the $B-L$ direction.

There is a further technical issue concerning the adjoint sector.
It is important that the adjoint Higgs transform non-trivially under
some flavor symmetry, otherwise it could be replaced by an explicit
$SO(10)$-singlet mass in the terms where it couples to quarks and leptons.
Consider, in particular, the second term in Eq. (3), ${\bf 16}_3 
\overline{{\bf 16}}\ {\bf 45}_H$, which is responsible
for the Georgi-Jarlskog factors of $\frac{1}{3}$ and $3$ that appear 
in the mass matrices shown in Eq. (1). If the ${\bf 45}_H$ has no
non-trivial flavor charges, then any flavor symmetry would also allow
the term $M {\bf 16}_3 \overline{{\bf 16}}$. This would replace the
factors of $3$ by a free parameter, so that the Georgi-Jarlskog factors
would no longer be predicted. On the other hand, any non-trivial 
superpotential for the adjoint Higgs is inconsistent with that field
having a $U(1)$ flavor charge. In the four-dimensional models of 
\cite{abb,ab,LMA}
this problem was solved by making ${\bf 45}_H$ negative under a $Z_2$
flavor symmetry, which is consistent with the form given in Eq. (5). 
However, such a symmetry would not allow the important cubic term in
Eq. (12). An interesting solution to this problem is made possible by
the brane structure, namely to introduce a singlet Higgs
field $W$, that lives only on the brane at $O'$ and that also has
negative parity under the $Z_2$ flavor symmetry. 
Then instead of the cubic term in Eq. (12) one
could have the term ${\rm tr}\ \Phi^3_{(15, 1, 1)} W/M$. Since the field $W$
lives only on the brane at $O'$, while all the quarks and leptons live on
the other brane, there is no way for $W$ to substitute for ${\bf 45}_H$
in quark and lepton Yukawa terms (which it would be allowed to do by
symmetries).

{\it The adjoint-spinor sector.}
In the four-dimensional version of the model it is crucial that
the VEV of the adjoint (because of its role in
the doublet-triplet splitting) point almost exactly in the $B-L$ direction.
Specifically, if $\langle {\bf 45}_H \rangle = A (B-L) + B I_{3R}$
then $B/A \leq 10^{-13}$. However, since in the five-dimensional version 
of the model the doublet-triplet splitting is done in another way,
which does not involve the adjoint Higgs field, there is no reason
why $B/A$ needs to be so small. The only role that $\langle {\bf 45}_H
\rangle \propto B-L$ plays is in explaining features of the quark and 
lepton masses and mixing, such as the Georgi-Jarlskog factors.
Thus, in the five-dimensional version of the model there is no harm in
there being deviations at the $10\%$ level of the adjoint VEV from
$B-L$. Consequently, the various operators (which we called ``Class B''
in the last section) that destabilize the Dimopoulos-Wilczek form are
not a danger in the five-dimensional model. One thing this means 
is that the coupling of the adjoint Higgs to the spinor Higgs need not
have the very special form shown in Eq. (7), but could in principle
have simpler and more obvious forms like 
$\overline{{\bf 16}}_H {\bf 45}_H {\bf 16}_H$.
However, in the specific five-dimensional realization of the model we
shall present below, the adjoint-spinor sector has the same form as 
in Eq. (7).

{\it The spinor sector.}
In the four-dimensional model we saw that the simple form for the 
spinor sector shown in Eq. (6) leads to problems. In particular,
the existence of those terms and of the term $({\bf 45}_H)^2$ in 
Eq. (5) implies
that no symmetry prevents terms of the form $\overline{{\bf 16}}_H
{\bf 16}_H ({\bf 45}_H)^2$, which destabilize the Dimopoulos-Wilczek form
of the adjoint VEV. Thus in \cite{ab} a more complicated
form for $W_{16}$ was chosen that involved the quartic term
$(\overline{{\bf 16}}_H {\bf 16}_H)^2$. Here, however, since we do not
have to worry about slightly deviating from the Dimopoulos-Wilczek form,
the spinor sector can have the simple form in Eq. (6), and indeed will
in the five-dimensional model that we will present.

We now present the rest of the details of the five-dimensional model.
It very closely parallels the four-dimensional model described
in \cite{abb,ab,LMA}, but with the simplifications already mentioned.
In the four-dimensional model a $U(1) \times Z_2 \times Z_2$ flavor symmetry 
was employed to control the form of the quark and lepton mass matrices and
prevent terms in the Higgs superpotential that would destabilize the gauge 
hierarchy. Here a slightly simpler $U(1) \times Z_2$ flavor symmetry 
is enough,
essentially because we do not need the doublet-triplet-splitting sector
of Eq. (4). Aside from the Higgs field $\tilde{\bf 10}_H$ which is absent
here, the fields that are needed are exactly those needed in
\cite{abb,ab,LMA}. The Higgs and matter superfields, along with the 
names previously used in the earlier papers, and their 
$U(1) \times Z_2$ flavor charges are listed in Tables I and II.

The Higgs superpotential (on the brane at $O$) contains the following
terms to be compared only with Eqs. (5)-(8), since there is no analogue 
of Eq. (4):

\begin{equation}
\begin{array}{rcl}
W_{Higgs} & = & -M_3 {\rm tr} ({\bf 45}_H)^2 \\ & & \\
& + & (\overline{{\bf 16}}_H {\bf 16}_H - M_4^2) \; X \\ & & \\
& + & \overline{{\bf 16}}_H ({\bf 45}_H \; P/M + P'){\bf 16}'_H 
 + \overline{{\bf 16}}'_H ({\bf 45}_H \ P/M + P'){\bf 16}_H  \\ & & \\
& + & \overline{{\bf 16}}_H \overline{{\bf 16}}_H {\bf 10}_H \; Y'/M 
 + {\bf 16}_H {\bf 16}'_H {\bf 10}_{2H} 
+ {\bf 10}_{1H} {\bf 10}_{2H} \; S_1. \\
\end{array}
\end{equation}

\noindent
The gauge hierarchy would be endangered by the terms we called class A before,
namely those like $M ({\bf 10}_H)^2$. However, these are eliminated by
the $U(1)$ flavor symmetry, just as in \cite{abb,ab,LMA}.
The terms that we called class B, which destabilize the minimum
$\langle {\bf 45}_H \rangle \propto B-L$, are not all excluded. However,
that is not important for the gauge hierarchy, since the doublet-triplet
splitting is not done here by the adjoint Higgs via the Dimopoulos-Wilczek
mechanism. Terms that do tend to push $\langle {\bf 45}_H \rangle$
away from the $B-L$ direction are 
$\overline{{\bf 16}}_H {\bf 45}_H {\bf 16}'_H \; P/M +
\overline{{\bf 16}}'_H  {\bf 45}_H {\bf 16}_H \; P/M$. However, if the 
coefficients of these terms are somewhat small, say a tenth, then the
VEV of the adjoint is still close enough to the $B-L$ direction to
give satisfactry Georgi-Jarlskog factors.

The nature of the massless Higgs doublets at the GUT scale can be determined 
from the $5 \times 5$ mass matrix for the Higgs multiplets which is easily 
constructed from the above superpotential. 

\begin{equation}
\left( \overline{{\bf 5}}_{10_H}, \overline{{\bf 5}}_{10_{1H}},
\overline{{\bf 5}}_{10_{2H}}, \overline{{\bf 5}}_{16_H},
\overline{{\bf 5}}_{16'_H}
\right) \left( 
\begin{array}{ccccc}
0 & 0 & 0 & y' & 0 \\
0 & 0 & s_1 & 0 & 0 \\
0 & s_1 & 0 & 0 & 0 \\
0 & 0 & 0 & x & p' \\
0 & 0 & c & p' & 0 
\end{array}
\right) \left(
\begin{array}{c}
{\bf 5}_{10_H} \\ {\bf 5}_{10_{1H}} \\ {\bf 5}_{10_{2H}} \\ 
{\bf 5}_{\overline{16}_H} 
\\ {\bf 5}_{\overline{16}'_H} 
\end{array} \right),
\end{equation} 

\noindent
where $y' = \langle \overline{{\bf 16}}_H Y'/M \rangle$, 
$s_1 = \langle S_1 \rangle$, $x = \langle X \rangle$, and $p' = 
\langle {\bf 45}_H P/M + P' \rangle$, and $c = \langle {\bf 16}_H \rangle$.
This matrix involves only the 
brane fields and the zero modes of the bulk fields, not the bulk modes
with superlarge K-K masses. One finds the Higgs doublet 
contributing to the up-type quark and Dirac neutrino mass matrices lies 
solely in the ${\bf 10}_H$ representation, i.e., it comes from the K-K zero
mode of that field:

\begin{equation}
	H_u \subset {\bf 5}_{{\bf 10}_H}, 
\end{equation}

\noindent while the Higgs doublet contributing to the down-type quark and 
charged lepton mass matrices consists of the linear combination 

\begin{equation}
	H_d \subset \overline{{\bf 5}}_{{\bf 10}_H} \cos \gamma 
		+ \overline{{\bf 5}}_{{\bf 16}'_H} \sin \gamma \cos \gamma'
		+ \overline{{\bf 5}}_{{\bf 10}_{1H}} \sin \gamma \sin \gamma'.
\end{equation}

\noindent
This interesting feature of $H_d$ allows the possibility of Yukawa coupling
unification with a moderate value of $\tan \beta \equiv v_u/v_d \sim 5$, 
provided $\gamma \sim \pi/2$ and $\gamma' \sim 0$.  Note also that the above 
equations reveal why ${\bf 10}_{1H}$ can develop a weak-scale VEV only in the 
$(1,2,-\frac{1}{2})$ direction as previously assumed.  

The light doublets given in Eqs. (15) and (16) have $SU(5)$ partners that
are color triplets. These color-triplet partners do not have a superlarge mass
term connecting them to each other as they would have to have in a 
four-dimensional model to make them superheavy. That is why, as we shall see,
there is no problem with proton decay mediated by exchange of these colored 
states. However, these triplets are superheavy because of the K-K masses 
(there are no K-K zero modes for the triplets). 

The Yukawa superpotential on the brane at $O$ for the charged quarks and 
leptons has the following form (which should be compared to 
Eq. (3)):

\begin{equation}
\begin{array}{ccl}
W_{Yukawa} & = & {\bf 16}_3 {\bf 16}_3 {\bf 10}_H + 
{\bf 16}_3 \overline{{\bf 16}}\; {\bf 45}_H + {\bf 16}_2 {\bf 16} \; {\bf 10}_H
+ {\bf 16} \overline{{\bf 16}} \; \; P' \\ & & \\
& + & {\bf 16}_3 {\bf 10}_2  {\bf 16}'_H +
{\bf 16}_2 {\bf 10}_1  {\bf 16}_H + {\bf 10}_1 {\bf 10}_2 Y \\ & & \\
& + & {\bf 16}_1 {\bf 16}_2 {\bf 10}_{1H} \; S_2/M + {\bf 16}_1 {\bf 16}_3 
  {\bf 10}_{1H} \; S_3/M + {\bf 16}' {\bf 16}' {\bf 10}_H \\ & & \\
& + & {\bf 16}_1 \overline{{\bf 16}}' \; Y' + {\bf 16}' \overline{{\bf 16}}' 
  \; S_4 + {\bf 16}_3 {\bf 1}_3 \overline{\bf 16}_H + {\bf 16}_2 {\bf 45}_2 
  \overline{\bf 16}_H \\
& & \\
& + & {\bf 45}_1 {\bf 45}_2 P' + {\bf 45}_1 {\bf 1}_3 {\bf 45}_H 
+ {\bf 1}_3 {\bf 1}^c_3 P + {\bf 1}^c_3 {\bf 1}^c_3 V_M, \\
\end{array}
\end{equation}

\noindent
where we have omitted additional terms and matter fields which contribute 
to the Dirac 11 mass matrix elements and the first row and first column
of the right-handed Majorana mass matrix.
Note that since there is a local $SO(10)$ symmetry on the brane at $O$,
the Yukawa superpotential has a full $SO(10)$ invariance.  As in 
\cite{ab}, the 2-3 sector of the Majorana mass matrix arises from the 
last six terms in the Yukawa superpotential, where the VEV of $V_M$ 
violates lepton number by two units.  In fact, these terms involving  
adjoint and singlet representations of the quarks and leptons generate
a Majorana mass matrix for the right-handed neutrinos that leads to 
the LMA solution of the solar neutrino problem via the seesaw mechanism.  

Let us now look at the question of proton decay from dimension-five 
operators more closely. In four dimensions, in order to make all color-triplet
Higgsinos superheavy, it is necessary to have mass terms in the 
superpotential that couple the $\overline{{\bf 3}}$ to the ${\bf 3}$ 
Higgsinos. These mass terms allow proton decay via dimension-five operators. 
In the 5d model, however, some of these mass terms can be dispensed with, 
since the triplets can get superlarge Kaluza-Klein masses, and thus proton 
decay from dimension-five operators can be avoided. That dimension-five 
proton decay operators are not a problem in this model can be seen in the 
following simple way. The mass matrix shown in Eq. (14) respects an approximate 
global symmetry. Under this symmetry the fields ${\bf 10}_H$, ${\bf 10}_{1H}$,
${\bf 16}_H$, and ${\bf 16}'_H$ all have charge $+1$; and the the fields 
${\bf 10}_{2H}$, $\overline{{\bf 16}}_H$, and $\overline{{\bf 16}}'_H$ all 
have charge $-1$. The only Higgs(ino) fields that couple to a pair of
light matter (i.e. quark/lepton) fields are those with global charge $+1$, 
as can be seen from the Yukawa part of the superpotential. Thus, the 
color $\overline{{\bf 3}}$ and ${\bf 3}$ Higgsinos that couple to light quarks
and leptons all have charge $+1$, and any mass term that coupled such
$\overline{{\bf 3}}$ and ${\bf 3}$ Higgsinos to each other would violate 
the global charge by two units, and would thus be forbidden. However, the 
global symmetry is not respected by certain high-dimension operators that 
contribute at higher order in $1/M_{Pl}$ to the mass matrix in Eq. (14). 
The most significant such operator is 
$({\bf 10}_{1H})^2 \langle (S_2)^2 V_M \rangle/M_{Pl}^2$. This term
allows a dimension-five proton decay operator, but one that is suppressed by
a factor of $O(M_{GUT}^2/M_{Pl}^2)$ and is therefore harmless.

\section{Conclusions}

We have taken a very complete $SO(10)$ model that exists in the literature
and shown that it can be transcribed, as it were, to five space-time
dimensions in such a way as to preserve its successful features 
(especially its predictions of quark and lepton masses and mixings)
but obviate the problem of proton decay mediated by colored Higgsino exchange,
which tends to give too rapid a decay in four-dimensional SUSY GUTs.

The idea is to put all the quarks and leptons and most of the Higgs
multiplets on a brane where there is the full local $SO(10)$, so that the
structure of the original four-dimensional $SO(10)$ model is left essentially 
intact, including all the predictions for quark and lepton masses
and mixings that rely on the group theoretical constraints imposed
on the superpotential by $SO(10)$. But the Higgs doublets of the MSSM come
from hypermultiplets in the five-dimensional bulk. This permits the 
doublet-triplet splitting
to be done in such a way as to prevent the dangerous proton decay
mediated by the exchange of colored Higgsinos, as has been shown in many
papers. To do this, $SO(10)$ is broken by the orbifold compactification down 
to the Pati-Salam group, with the further breaking down to the Standard Model
group done by the conventional four-dimensional Higgs mechanism on the 
physical $SO(10)$ brane as suggested by Derm\'{i}\v{s}ek and 
Mafi \cite{DM}.

Because the doublet-triplet splitting is no longer done by the
adjoint Higgs field, as in the original four-dimensional model,
it is not important in the five-dimensional model that the
``Dimopoulos-Wilczek form'' of the adjoint VEV be protected to extremely
high accuracy to preserve the gauge hierarchy. This relaxes many
of the conditions on the Higgs sector that had to be met in the 
four-dimensional model to keep the gauge hierarchy natural, and thus
allows some simplification of the Higgs sector. Since the adjoint Higgs
in the four-dimensional model played a key role in giving a realistic
pattern of quark and lepton masses, it must still be present in the
five-dimensional version, even though it no longer plays a role in
doublet-triplet splitting. By putting the adjoint in the bulk, 
its VEV can be driven to the desired $B-L$ direction by superpotential
terms on the hidden Pati-Salam brane. 

The approach described in this paper should be applicable to virtually
all realistic four-dimensional $SO(10)$ models. Thus, it provides a 
way of curing any such model of problems with proton decay coming from 
dimension-five operators without affecting its successful features 
or its predictions for quark and lepton masses and mixings.

\vspace*{0.3in}

The research of SMB was supported in part by Department of Energy Grant
Number DE FG02 91 ER 40626 A007.  One of us (CHA) thanks the Fermilab
Theoretical Physics Department for its kind hospitality where his work
was carried out.  Fermilab is operated by Universities Research Association
Inc. under contract with the Department of Energy.

\newpage
\hspace*{0.5in}Table I.\ \begin{minipage}[t]{4.5in}{Higgs 
	superfields and their $SO(10)$ and $U(1) \times Z_2$
	flavor transformation assignments.}\end{minipage}\\[0.2in]
$$\begin{array}{|c|c|c|c|c|}
\hline & & & &\\
 & {\rm \quad Previous\quad }  & \quad SO(10)\quad  &\quad U(1)\quad & \ Z_2\ \\
 & {\rm Label} & & &\\
& & & &\\\hline
\quad {\bf 45}_H\quad  & A & {\bf 45} & 0 & - \\
{\bf 16}_H & C & {\bf 16} & \frac{3}{2} & - \\
\overline{{\bf 16}}_H & \overline{C} & \overline{{\bf 16}} & -\frac{3}{2} & - \\
{\bf 16}'_H & C' & {\bf 16} & \frac{3}{2} - p & + \\
\overline{{\bf 16}}'_H & \overline{C}' & \overline{{\bf 16}} & 
	-\frac{3}{2} - p & + \\
{\bf 10}_H & T_1 & {\bf 10} & 1 & + \\
{\bf 10}_{1H} & T'_0 & {\bf 10} & -2-q & + \\
{\bf 10}_{2H} & \bar{T}_0 & {\bf 10} & -3+p & - \\
X & X & {\bf 1} & 0 & + \\
P^{(\prime)} & P & {\bf 1} & p & +(-) \\
Y^{(\prime)} & Y & {\bf 1} & 2 & -(+) \\
S_1 & S' & {\bf 1} & 5-p+q & - \\
S_2 &  & {\bf 1} & q-p & + \\
S_3 &  & {\bf 1} & q & + \\
S_4 & S & {\bf 1} & 5 & - \\
V_M & V_M & {\bf 1} & 4+2p & + \\\hline
\end{array}$$

\vspace*{0.25in}
\hspace*{0.5in}Table II.\ \begin{minipage}[t]{4.5in}{Matter superfields
	and their $SO(10)$ and $U(1) \times Z_2$ flavor transformation 
	assignments.}\end{minipage}\\[0.2in]
$$\begin{array}{|c|c|c|c|}
\hline & & &\\
 & \qquad SO(10)\qquad & \qquad U(1) \qquad & \ Z_2\ \\
& & &\\ \hline
\qquad {\bf 16}_1\qquad  & {\bf 16} & \frac{5}{2} & - \\
{\bf 16}_2 & {\bf 16} & -\frac{1}{2} +p & - \\
{\bf 16}_3 & {\bf 16} & -\frac{1}{2} & - \\
{\bf 16} & {\bf 16} & -\frac{1}{2} - p & - \\
\overline{\bf 16} & \overline{{\bf 16}} & \frac{1}{2} & + \\
{\bf 16}' & {\bf 16} & -\frac{1}{2} & + \\
\overline{\bf 16}' & \overline{{\bf 16}} & -\frac{9}{2} & - \\
{\bf 10}_1 & {\bf 10} & -1-p & + \\
{\bf 10}_2 & {\bf 10} & -1+p & - \\
{\bf 1}_3 & {\bf 1} & 2 & + \\
{\bf 1}_3^c & {\bf 1} & -2-p & + \\
{\bf 45}_1 & {\bf 45} & -2 & - \\
{\bf 45}_2 & {\bf 45} & 2-p & + \\\hline
\end{array}$$
\end{document}